# $SU(3)_C \otimes SU(3)_L \otimes U(1)_X$ gauge symmetry from $SU(4)_{PS} \otimes SU(4)_{L+R}$

Sutapa Sen and Aparna Dixit

Department of Physics, Christ Church P.G. College, Kanpur - 208001, India

Abstract

We consider an extension of the standard model gauge symmetry to a local gauge group $SU(3)_C \otimes SU(3)_L \otimes U(1)_X$ which is a subgroup of $SU(4)_{PS} \otimes SU(4)_{L+R}$ .The symmetry breaking pattern is $SU(4) \rightarrow SU(3) \otimes U(1)$ for both weak $SU(4)_{L+R}$ and strong Pati-Salam $SU(4)_{PS}$ group. The $SU(3)_C \otimes U(1)_{B-L} \otimes SU(3)_L \otimes U(1)_{Y1}$ (3-3-1-1) local gauge symmetry breaks to $SU(3)_C \otimes SU(3)_L \otimes U(1)_X$ and generates a 3-3-1 model with three-generation, anomaly-free fermions which transform as bifundamentals of (3-3-1-1).

The 3-3-1 model is of Pleitez-Frampton type but $SU(3)_L$ gauge bosons (B-L = 0) do not include bilepton gauge boson.The neutral gauge bosons include $\gamma$, Z, $Z'$ and a fourth, heavy gauge boson $Z''$ which decouples from rest but can decay to ordinary fermions. An analysis for two -body decays of neutral $Z'$ gauge boson is presented .From Yukawa interactions, the masses of all exotic fermions are obtained in the TeV region .This restricts $Z'$ decays to exotic fermions The $Z'$ is also found to be leptophobic and decays mainly to quarks.

## 1. Introduction

One of the simplest extension of the Standard model is to enhance the $SU(2) \otimes U(1)$ symmetry to SU(3) and SU(4) groups. This provides quark - lepton unification at TeV scale and the weak mixing angle $\theta_W$ as[1] $\sin^2 \theta_W = 0.25$ at the unification scale. While SU(3) and SU(4) gauge symmetry work only for leptons , the symmetry has to be enhanced to $SU(3) \otimes U(1)_X$ to accommodate fractionally charged quarks as well [2] .The X charges are fixed by anomaly cancellations within three generations of fermions.The $SU(3)_C \otimes SU(3)_L \otimes U(1)_X$ symmetry include cases of subgroups of higher symmetry group as $E_6$ or $SU(6)\otimes U(1)$ [3] with gauge bosons having ordinary electric charges .

A second type of 3-3-1 models include quarks with exotic electric charges – 4/3 and 5/3 and doubly charged gauge bosons [2]. The electric charge operator Q is a linear combination of three generators of the $SU(3)_L$ group,

$\frac{Q}{e} = T_{3L} + a\, T_{8L} + XI_3$ , where $T_{iL} = \frac{1}{2} \lambda_{iL}$ , $i = 1,2,..8$ ; $\lambda_{iL}$ are Gell-Mann matrices for SU(3) group with normalization $Tr(\lambda_{iL}\lambda_{jL}) = 2\,\delta_{ij}$ and $I_3 = Dg\,(1,1,1)$ unit matrix. For models with no doubly charged gauge bosons [3], $a = -1/\sqrt{3}$ while $a = \pm\sqrt{3}$ for the class of models with doubly charged bosons [4]. An extension of local gauge group to $SU(3)_C \otimes SU(4)_L \otimes U(1)_X$ has been recently proposed for three-family models with gauge bosons without exotic charges [5]The Little Higgs models also deal with extended electroweak gauge symmetry 3-3-1 and 3-4-1 around TeV scale [6] A general expression for electric charge operator in the 3-4-1 models is [5]

$$\frac{Q}{e} = T_{3L} + a\, T_{8L} + b\, T_{15L} + X I_4, \qquad (1)$$

where $T_{iL} = \frac{1}{2}\lambda_{iL}$, i = 1,2,..8 ; $\lambda_{iL}$ are Gell-Mann matrices for SU(4) group with normalization $Tr(\lambda_{iL}\lambda_{jL}) = 2\delta_{ij}$ and $I_4 = Dg\,(1,1,1,1)$ unit matrix. For models with gauge bosons with electric charges 0 , $\pm$ 1 only, the possibilities for simultaneous values of a , b are [5]

$$a = b = \pm\, 1/\sqrt{3};\quad a = 1/\sqrt{3},\ b = -2/\sqrt{3}\ ;\quad a = -1/\sqrt{3},\ b = 2/\sqrt{3}$$

In this work we consider a version of 3-3-1 model with doubly charged gauge boson and charged heavy leptons as derived from a 3-3-1-1 gauge symmetry which is a subgroup of $SU(4)_{PS} \otimes SU(4)_{L+R}$. This group has been recently considered in 5D supersymmetric case [7] as a subgroup of SO(12) and $E_7$ .The $SU(4)_{PS}$ strong gauge group as well as $SU(4)_{L+R}$ flavor group are decomposed as $SU(4) \supset SU(3) \otimes U(1)$ , such that local gauge symmetry is

$$G = SU(3)_C \otimes SU(3)_L \otimes U(1)_{B-L} \otimes U(1)_{Y1} \qquad (2)$$

The SU(4) generators $\frac{B-L}{2} = Dg\,(1/6,1/6,1/6,-1/2) = \sqrt{2/3}\; T_{15\,PS}$ ;

$$Y_1 = Dg\,(1/2,\, 1/2,\, 1/2,\, -3/2) = \sqrt{6}\, T_{15\ L+R}$$

The electric charge operator is

$$\frac{Q}{e} = T_3 - \sqrt{3}\, T_8 + \sqrt{6}\, T_{15} + \frac{(B-L)}{2} I_4 \qquad (3)$$

where $T_\alpha$ , $\alpha$ = 1,2,.15 are diagonal generators of $SU(4)_{L+R}$ and $I_4$ is the 4 x 4 identity matrix.

$$T_3 = \frac{1}{2} Dg\,(1,-1,0,0),\quad T_8 = \frac{1}{2\sqrt{3}} Dg\,(1,1,-2,0) \quad \text{and} \quad T_{15} = \frac{1}{2\sqrt{6}} Dg(1,1,1,-3)$$

By extending the 3-3-1 gauge symmetry to 3-3-1-1 which is embedded in $SU(4)_{PS} \otimes SU(4)_{L+R}$ gauge group, a bifundamental pattern is obtained for fermions and scalars. An important point of difference of our model from all other versions of 3-3-1 models [4 ] is that (B-L) baryon minus lepton number is conserved and is an operator of the strong $SU(4)_{PS}$ group. The gauge bosons of the model include singly and doubly charged vector bosons which are not bileptons and have B-L = 0.The doubly charged exotic gauge boson can couple only to exotic quarks and lepton .The neutral bosons include photon field $\gamma$, Z, $Z^{/}$ and a fourth boson $Z^{//}$ which decouples from the rest. The fourth gauge boson couples to ordinary fermions and has a large decay width and mass.

The $SU(3)_C \otimes SU(3)_L \otimes U(1)_X$ gauge symmetry is obtained due to symmetry breaking $U(1)_{Y1} \otimes U(1)_{B-L} \rightarrow U(1)_X$ at some scale $M_U$ by Higgs mechanism with a new neutral scalar $\Phi_0$ with lepton number L = -3, which has vacuum expectation value aligned along $< \Phi_0> = V^{/}$. In section 2 we present a general formalism of the model and consequences of $SU(4)_C \otimes SU(4)_{L+R}$ symmetry breaking. Section 3 deals with the gauge boson sector with mixing and mass of neutral gauge bosons. Section 4 deals with the fermion content of the model and scalars needed for symmetry breaking. In section 5, the charged and neutral currents along with Yukawa couplings and fermion masses and mixing are discussed. The phenomenology of the Z, $Z^{/}$ ,$Z^{//}$ gauge bosons and possible decays to quarks, leptons and scalars are presented in section 6. Section 7 is a brief discussion of results and conclusions.

## 2. **General Formulation**

Recently, the $SO(12) \supset SU(4)_C \otimes SU(4)_{L+R}$ has been considered for grand unification group [7] in a 5D formalism. While $SU(4)_{PS} \supset SU(3)_C \otimes U(1)_{B-L}$ breaking mechanism by (15,1) Higgs at a scale $M_U$ [11] gives leptoquark gauge bosons at high energies, we do not consider these bosons and assume a 3-4-1 symmetry.

The decomposition of $SU(4)_{L+R} \supset SU(2)_L \otimes SU(2)_R \otimes U(1)_{Y1}$ in the left-right symmetric model gives [7]

$$4_{L+R} = (2_L, 1_R, 1/2) + (1_L, 2_R, -1/2) = (x_u, x_d) + (y_u, y_d) \quad (4)$$

where $Y_1$ is the SU(4) operator, $Y_1 = \frac{1}{2\sqrt{2}} Dg(1,1,-1,-1)$

We consider the decomposition $SU(4)_{L+R} \supset SU(3)_L \otimes U(1)_{Y1}$ with

$$4_{L+R} = (3_L, 1/2) + (1_R, -3/2) = (x_u, x_d, \bar{y}_d\, \bar{y}_d\, y_u) + (y_d\, y_d\, y_d)$$

where the SU(4) operator $Y_1 = \sqrt{6}\, T_{15\ L+R}$ (5)

The gauge bosons belong to 15-plet representation of SU(4) and can be decomposed as

$$\underline{15} = (\underline{8}, 0) + (\underline{3}, 2) + (\underline{3^*}, -2) + (\underline{1}, 0) \quad (6)$$

The flavor $SU(4)_{L+R}$ decomposition gives $8_L$ $SU(3)_L$ gauge bosons, (3 + 3*) X bosons and a neutral singlet $Y_1^{\mu}$ associated with $U(1)_{Y1}$. The strong $SU(4)_{PS}$ group is decomposed as $SU(4)_{PS} \rightarrow SU(3)_C \otimes U(1)_{B-L}$ where $SU(3)_C$ is the color QCD gauge group and operator $\frac{(B-L)}{2} = Dg\left(\frac{1}{6}, \frac{1}{6}\ \frac{1}{6}, -\frac{1}{2}\right) = \frac{2}{\sqrt{6}} T_{15\,PS}$ .

$$\underline{4}_{PS} = (\underline{3}_C, 1/6) + (\underline{1}, -1/2) = x_C + x_S \quad (7)$$

The gauge bosons include 8 gluons , 6 ( 3 + 3*) leptoquarks and a neutral singlet $B^{\mu}(1,0)$ associated with $U(1)_{B-L}$.

The $SU(4)_{L+R}$ gauge bosons belong to the 15 –plet adjoint representation

$$\lambda_m W^{\mu m} = \begin{pmatrix} D_1^{\mu} & \sqrt{2}\, W^{+\mu} & \sqrt{2} Y^{-\mu} & \sqrt{2} X^{++} \\ \sqrt{2} W^{-\mu} & D_2^{\mu} & \sqrt{2} Y^{--\mu} & \sqrt{2} X^{+\mu} \\ \sqrt{2} Y^{+\mu} & \sqrt{2} Y^{++\mu} & D_3^{\mu} & \sqrt{2} X^{+++\mu} \\ \sqrt{2} X^{--\mu} & \sqrt{2} X^{-\mu} & \sqrt{2} X^{---\mu} & D_4^{\mu} \end{pmatrix}$$

where m = 1,2,….15 and

$D_1^{\mu} = W^{3\mu} + W_8^{\mu}/\sqrt{3} + 1/\sqrt{6}\ W_{15}^{\mu}$ ; $D_2^{\mu} = -W_3^{\mu} + W_8^{\mu}/\sqrt{3} + W_{15}^{\mu}/\sqrt{6}$;

$D_3^{\mu} = -2/\sqrt{3}\ W_8^{\mu} + W_{15}^{\mu}/\sqrt{6}$ ; $D_4^{\mu} = -3/\sqrt{6}\ W_{15}^{\mu}$ .

The $SU(4)_{L+R}$ → $SU(3)_L \otimes U(1)_{Y1}$ breaking by 15-plet Higgs scalar [11] is assumed so that the local gauge symmetry group

$$G = SU(3)_C \otimes U(1)_{B-L} \otimes SU(3)_L \otimes U(1)_{Y1} = G_C \otimes G_L$$

This is broken at unification scale as $U(1)_{Y1} \otimes U(1)_{B-L} \rightarrow U(1)_X$ , where

$$X = Y_1 + \frac{(B-L)}{2} = \sqrt{6}\, T_{15} + \frac{(B-L)}{2} I_4 \qquad (8)$$

A neutral gauge boson $V^\mu$ is associated with $U(1)_X$. We consider a neutral scalar

$\Phi_0 = (1_C, 1_L, -3/2, 3/2) = (y_d\, y_d\, y_d\, \bar{x}_S\, \bar{x}_S\, \bar{x}_S) = (1_C, 1_L, 0)$ aligned with vacuum expectation value [VEV} $<\Phi_0> = V^{/} I_4$ The physical gauge fields $V^\mu$ and a fourth neutral boson $Z^{//\mu}$ are defined as

$$B^\mu = \cos\theta\, V^\mu - \sin\theta\, Z^{//\mu}; \quad Y_1^{\ \mu} = \sin\theta\, V^\mu + \cos\theta\, Z^{//\mu} \qquad (9)$$

The unification condition gives

$$\frac{!}{g_X^{\ 2}} = \frac{1}{g_{B-L}^{\ 2}} + \frac{1}{g_{15}^{\ 2}}; \quad \tan\theta = g_{B-L} / g_{15} \ ; g_X/g_{15} = \sin\theta$$

The covariant derivative

$$D^\mu = \partial^\mu - i\, g\, T_a W^{a\,\mu} - i\, g_X X\, V^\mu - i\, g\, T^b W_b^{\ \mu} - \frac{i g_X}{\sin\theta \cos\theta} (\sqrt{6}\, T_{15} - X \sin^2\theta) Z^{//\mu}$$

where a = 1,2,..8 and b = 9,10,….14. (10)

The neutral gauge boson $Z^{//\mu}$ decouples from the rest and acquires a large mass,

$$M^2_{Z^{//}} = \frac{9\, g_X^{\ 2} V^{/2}}{2 \sin^2\theta \cos^2\theta}$$

The pattern of symmetry breaking in the present case is

$$SU(3)_C \times SU(3)_L \times U(1)_{Y1} \times U(1)_{B-L} \xrightarrow{M_U} SU(3)_C \times SU(3)_L \times U(1)_X$$

$$SU(3)_L \otimes U(1)_X \xrightarrow{M_\chi} SU(2)_L \otimes U(1)_X^{/} \otimes U(1)_X \xrightarrow{M_\chi} SU(2)_L \otimes U(1)_Y \xrightarrow{M_\eta, M_\rho} U(1)_{em}$$

Further breaking of $SU(3)_L \otimes U(1)_X$ symmetry requires three scalar triplets

$\chi, \rho, \eta$ with VEV aligned in the directions

$$<\chi^0> = [0, 0, \frac{V}{\sqrt{2}}]^T; \quad < \rho^0> = [0, \frac{u}{\sqrt{2}}, 0 ]^T; \quad <\eta^0> = [ \frac{v}{\sqrt{2}}, 0, 0 ]^T \qquad (12)$$

$$\eta\,(1,3,0) = (\eta^0,\eta_1^-, \eta_2^+)^T; \rho(1,3\,,1) = ( \rho^+, \rho^0, \rho^{++})^T; \; \chi\,(1,3,-1) = (\chi^-, \chi^{--}, \chi^0)^T \qquad (13)$$

The sextet scalar is not required for charged lepton masses in this case.

We consider [12] spontaneous symmetry breaking $U(1)_{X'} \otimes U(1)_X \rightarrow U(1)_Y$ by Higgs scalar $\chi$ ,where $X' = \sqrt{3}\, T_{8L}$, and $g_X' = g/\sqrt{3}$ . The unification condition gives

$$\frac{1}{g_Y^2} = \frac{1}{g_X^2} + \frac{1}{g_X'^2} \; ; \; g_Y^2 = \frac{g^2 g_X^2}{(3g_X^2 + g^2)} \; ; \; \tan\theta_W = \frac{g_Y}{g} = \frac{g_X}{\sqrt{(3g_X^2 + g^2)}}$$

$$\frac{g_X}{g} = \frac{\sin\theta_W}{\sqrt{( 1 - 4 \sin^2\theta_W)}} \qquad (14)$$

The weak mixing angle $\sin^2\theta_W \leq 0.25$ at unification scale. An important difference from 3-3-1 model with orbifold breaking [12] lies in the hypercharge operator Y defined as $Y = 2 ( X - X' )$.

**Case (1): a = $-\sqrt{3}$, b = $\sqrt{6}$**

From eqn (8) the electric charge operator

$$\frac{Q}{e} = T_{3L} + \frac{Y}{2} = T_{3L} - \sqrt{3}\, T_{8L} + \sqrt{6} T_{15} + \frac{(B-L) I_4}{2} \qquad (15)$$

In terms of SU(4) operator $T_{3R} = \frac{1}{2} Dg ( 0,0, 1, - 1) = \sqrt{3}\, T_{8L} - \sqrt{6} T_{15}$

$Q = T_{3L} - T_{3R} + \frac{(B-L)}{2} I_4$ which corresponds to the flipped model [10]

**Case (2) a = $\sqrt{3}$, b = $-\sqrt{6}$**

A second type of 3-3-1 model with a = $\sqrt{3}$ exists in literature [4,13] with mathematically equivalent formalism .This is obtained in the 3-3-1-1 model by choosing an

- Orthogonal combination for vector boson for $V^\mu$ associated with $U(1)_X$ and
- Redefining hypercharge Y.

We consider $U(1)_{Y1} \otimes U(1)_{B-L} \rightarrow U(1)_X$ breaking with charge X defined as the orthogonal combination $X = -Y_1 + \frac{(B-L)}{2} = -\sqrt{6}\, T_{15} + \frac{(B-L)}{2} I_4$

The hyper charge operator is defined as $\frac{Y}{2} = X + X^{/}$ [12] while the charge operator

$$\frac{Q}{e} = T_{3L} + \sqrt{3}\, T_{8L} - \sqrt{6}\, T_{15} + \frac{(B-L)\, I_4}{2}$$

In terms of SU(4) operator $T_{3R}$, $Q = T_{3L} + T_{3R} + \frac{(B-L)}{2}$, which corresponds to Georgi-Glashow model [9].The orthogonal combination for $Z^{//\mu}$ is changed in this case while $3_L \rightarrow 3_L^*$ adjoint representation for quarks and leptons. This requires a conjugate adjoint representation for gauge bosons .This case has been considered in literature [13]. We shall consider only case (1) in this work .

**3. Gauge boson sector and masses in 3-3-1 model.**

The charged gauge bosons acquires masses as given below

$$M^2(Y^{\pm}) = \tfrac{1}{2} g^2 (V^2 + v^2),\ M^2(Y^{\pm\pm}) = \tfrac{1}{2} g^2 (V^2 + u^2);\ M^2(W^{\pm}) = \tfrac{1}{2} g^2 (v^2 + u^2) \qquad (16)$$

The neutral gauge bosons include the photon field $A^\mu$ , massive physical fields $Z^\mu$, $Z^{/\mu}$ and a fourth heavy gauge boson $Z^{//\mu}$ so that the covariant derivative

$$D^\mu = \partial^\mu - i\, e\, Q A^\mu - \frac{i\, e}{s_W c_W} ( T_{3L} - Q\, s_W^2)\, Z^\mu + ig \frac{\sqrt{(1-4\, s_W^2)}}{c_W} \left[T_{8L} + \frac{\sqrt{3}\, s_W^2\, X}{(1-4\, s_W^2)}\right] Z^{/\mu}$$

$$- \frac{i\, g_X}{\sin\theta \cos\theta} [\sqrt{6} T_{15} - \sin^2\theta X] Z^{//\mu} - \frac{i\, g}{\sqrt{2}} [\lambda_{12} W^{\mu +} + \lambda_{45} Y^{\mu -} + \lambda_{67} Y^{\mu --} + H.c] \qquad (17)$$

*where* $\lambda_{ab} = \tfrac{1}{2} (\lambda_a + i\lambda_b) = T_a + iT_b$ and $\sin\theta_W = s_W$ *(eqn.14)*

The photon field $A^\mu$ and the fields $Z^\mu$ and $Z^{/\mu}$ are given by

$$A^\mu = s_W W_3^\mu + c_W [\sqrt{3}\, t_W W_8 - \sqrt{(1-3\, t_W^2)}\, V^\mu]$$

$$Z^\mu = c_W W_3^\mu - s_W [\sqrt{3}\, t_W W_8^\mu - \sqrt{(1-3\, t_W^2)}\, V^\mu]$$

$$Z'^\mu = \sqrt{(1-3\, t_W^2)}\, W_8^\mu + \sqrt{3}\, t_W V^\mu \qquad (18)$$

The Y hypercharge associated with SM abelian gauge boson is

$$Y^{\mu} = [\sqrt{3}\, t_W\, W_8^{\mu} - \sqrt{(1-3\, t_W^2)}\, V^{\mu}] \qquad (19)$$

For the mass spectrum of neutral gauge bosons, we obtain

$$M^2 = g^2/4\, [V^2\{2W_8^{\mu}/\sqrt{3} + 2g_X/g\ V^{\mu}\}^2 + u^2\{(-W_3^{\mu} + W_8^{\mu}/\sqrt{3}) + 2g_X/g\, V^{\mu}\}^2$$
$$+ v^2\{W_3^{\mu} + W_8^{\mu}/\sqrt{3}\}^2] \qquad (20)$$

*where* $g_X/g = s_W/\sqrt{(1-4\, s_W^2)}$

This gives zero mass for photon while masses for $Z^{\mu}$ , $Z^{/\mu}$ are obtained from

$$M_Z^2 = \frac{g^2}{2c_W^2}(v^2 + u^2) \qquad (21)$$

$$M_{Z}^{/\,2} = \frac{g^2}{3}\left[\frac{V^2 c_W^2}{(1-4s_W^2)} + \frac{(1-4\,s_W^2)}{4c_W^2}(v^2+u^2) + \frac{3u^2 s_W^2}{(1-4s_W^2)}\right] \qquad (22)$$

$$M^2_{ZZ^/} = \frac{g^2\sqrt{(1-4s_W^2)}}{4\sqrt{3}c_W^2}\left[v^2 - 2u^2\left(\frac{1+2s_W^2}{(1-4\,s_W^2)}\right)\right] \qquad (23)$$

From the mixing of $Z^{\mu}$ , $Z^{/\mu}$ through angle φ, $\tan 2\varphi = 2M^2_{ZZ^/}/(M^2_{Z^/} - M^2_Z)$

$\varphi = M^2_{ZZ^/}/M^2_{Z^/}$.

The mixing between Z, $Z^/$ can be considered by defining

$$Z_1^{\mu} = \cos\varphi Z^{\mu} - \sin\varphi Z^{/\mu}\ ;\quad Z_2^{\mu} = \sin\varphi Z^{\mu} + \cos\varphi Z^{/\mu} \qquad (24)$$

$$M^2_{Z1,\,Z2} = \frac{1}{2}[M_Z^2 + M_Z^{/\,2} \pm \sqrt{\{(M_Z^2 + M_Z^{/\,2})^2 - 4(M_Z^2 M_Z^{/\,2} - \varphi^2 M_Z^{/\,4})\}}] \qquad (25)$$

**4. Fermions and Scalars .**

In the present model, bifundamental structures are obtained for fermions and scalars with six basic fundamentals

$G_C = SU(3)_C \otimes U(1)_{B-L}$ : M = $x_C$ = $(3_C, 1/6)$, N = $x_S = (1_C, -1/2)$, $N^/ = (1_C, -3/2)$

$G_L = SU(3)_L \otimes U(1)_{Y1}$ : a = $(3_L, 1/2)$; b = $y_d\, y_d\, y_d = (1_R, -3/2)$ ; c= $y_d = (1_R, -1/2)$

The X- charges are defined by $X = Y_1 + (B-L)/2$

Table 1 lists three generation , anomaly-free fermion representation in which the $SU(3)_L$ triplet $(3_L, 1/2)$ is considered for both quarks and leptons.The (8,0) gauge bosons have B-L = 0 and are not bileptons.

TABLE 1

Three – generations of anomaly-free fermions as bifundamentals of $3_C- 3_L - 1_{Y1} - 1_{B-L}$

| Fermion | content | 3-3-1-1 | 3-3-1 |
|---|---|---|---|
| $Q_i$ | $(d_{Li}, u_{Li}, D_{Li})$ | a*M | $(3_C, 3_L^*, -1/3)$ |
| $Q_3$ | $(t_L, b_L, T_L)$ | aM | $(3_C, 3_L, 2/3)$ |
| $u_{R\alpha}^{\ c}$ | | cM* | $(3_C^*, 1_R, -2/3)$ |
| $d_{R\alpha}^{\ c}$ | | c*M* | $(3_C^*, 1_R, 1/3)$ |
| $D_{Ri}^{\ c}$ | $(D_{R1}^{\ c}, D_{R2}^{\ c})$ | b* M* | $(3_C^*, 1_R, 4/3)$ |
| $T_R^{\ c}$ | | bM* | $(3_C^*, 1_R, -5/3)$ |
| $L_\alpha$ | $(\nu_\alpha, e_\alpha, P_\alpha^{\ +})$ | aN | $(1_C, 3_L, 0)$ |
| $e_{R\alpha}^{\ c}$ | | c* N* | $(1_C, 1_R, 1)$ |
| $P_{R\alpha}^{\ c}$ | | bN* | $(1_C, 1_R, -1)$ |

Here i = 1,2 and $\alpha$ = 1,2,3 denote family indices.

The scalars are also bifundamentals and include

| | | | |
|---|---|---|---|
| $\eta$ | $(\eta^0, \eta_1^{\ -}, \eta_2^{\ +})^T$ | a c | $(1_C, 3_L, 0)$ |
| $\rho$ | $(\rho^+, \rho^0, \rho^{++})^T$ | a c* | $(1_C, 3_L, 1)$ |
| $\chi$ | $(\chi^-, \chi^{--}, \chi^0)^T$ | a b | $(1_C, 3_L, -1)$ |
| $\Phi_0$ | | b N $^{/}$* | $(1_C, 1_L, 0)$ |

The above particles correspond to an anomaly-free, three generation Pleitez-Frampton model with heavy leptons and exotic quarks. The special assumption of the present model is that in $\underline{3}_L$ the third flavor is $T_{3R} = \pm 3/2$.The leptons$\Psi_L(\nu,e^-,P^+)$ are obtained as a result of the triplet $3_L$ structure and symmetry breaking pattern for the model. The positron $e_{R\alpha}{}^c$ ( 1,1,1) and heavy lepton $P_{R\alpha}{}^c$ (1,1,-1) have to be added in this case. The scalar sector also do not require a sextet scalar field and consists of three triplet scalars and one new scalar $\Phi_0$ with L = - 3. The mass of $\Phi_0$ is obtained from symmetry breaking as $m^2_{\Phi 0} \propto V^{/2}$ where $<\Phi_0> = V^/ >> V$. There is no further difference for scalar sector from detailed work by several authors [13,14]

We consider the relations between ordinary and physical scalars in eqn. (12),(13)

$$\rho^{++} = \frac{1}{\sqrt{(u^2 + V^2)}}[u\, G^{++} + V H^{++}] \;\; ; \;\; \chi^{++} = \frac{1}{\sqrt{(u^2 + V^2)}}[VG^{++} + u\, H^{++}]$$

$$\eta_1^{+} = \frac{1}{\sqrt{(u^2 + v^2)}}[-v\, G_1^{+} + uH_1^{+}] \;\; ; \;\; \rho^{+} = \frac{1}{\sqrt{(u^2 + v^2)}}[u\, G_1^{+} + v\, H_1^{+}]$$

$$\eta_2^{+} = \frac{1}{\sqrt{(V^2 + v^2)}}[-v\, G_2^{+} + uH_2^{+}] \;\; ; \;\; \chi^{+} = \frac{1}{\sqrt{(V^2 + v^2)}}[V\, G_2^{+} + vH_2^{+}] \qquad (26)$$

For neutral scalars , the vacuum expansion of $\eta^0, \rho^0, \chi^0$ are

$$\eta^0 = v + \xi_\eta + i\zeta_\eta \,;\, \rho^0 = u + \xi_\rho + i\zeta_\rho \,;\, \chi^0 = V + \xi_\chi + i\zeta_\chi$$

$$\xi_\eta = \frac{1}{\sqrt{(u^2 + v^2)}}(v\, H_1^0 + u\, H_2^0)\, ; \;\; \xi_\rho = \frac{1}{\sqrt{(u^2 + v^2)}}(u\, H_1^0 - vH_2^0)\, ;\, \xi_\chi = H_3^0 ;\; \zeta_\chi = h^0. \qquad (27)$$

## 5.(a) Neutral Currents

The neutral current Lagrangian contains an additional gauge boson $Z^{//}$

$$L_{NC} = -e\, J_{em\,\mu} A^{\mu} - g\, J(Z)_\mu Z^\mu - \frac{g}{c_W} J(Z^/)_\mu Z^{/\mu} - 2\frac{g\, s_W}{c_W\, \sin 2\theta \sqrt{(1- 4s_W^2)}} J(Z^{//})_\mu Z^{//\mu} \qquad (28)$$

The Lagrangian for Z, $Z^/$ couplings to fermions

$$L(Z) = \frac{g}{2c_W} \Sigma_f\, \bar{f}\, \gamma_\mu [(g_L P_L + g_R P_R)] f Z^\mu \;\; ; \;\; g_L = T_{3L} - Qs_W^2 ;\; g_R = -Qs_W^2$$

$$= \frac{g}{2c_W} \Sigma_f \bar{f} \gamma_\mu ( v_Z^f + a_Z^f \gamma_5) f Z^\mu \; ; \; v_Z^f = T_{3L} - 2Q s_W^2 , \quad a_Z^f = - T_{3L} \tag{29}$$

$$L(Z^/) = \frac{g}{c_W} \Sigma_f \bar{f} \gamma_\mu [( g^/_L P_L + g^/_R P_R)] f Z^{/\mu}$$

$$\text{where } g^/_L = \sqrt{(1-4s_W^2)}\, T_{8L} + \frac{\sqrt{3}\, s_W^2\, X}{\sqrt{(1-4s_W^2)}} ; \quad g^/_R = + \frac{\sqrt{3}\, s_W^2\, X}{\sqrt{(1-4s_W^2)}} ; \tag{30}$$

$$L(Z^/) = \frac{g}{2c_W} \Sigma_f \bar{f} \gamma_\mu ( v_Z^{/f} + a_Z^{/f} \gamma_5) f Z^{/\mu} ,$$

$$\text{where } v_Z^/ = g^/_L + g^/_R \; ; a_Z^/ = - g^/_L + g^/_R \tag{31}$$

$$v_Z^{/f} = \sqrt{(1-4s_W^2)}\, T_{8L} + 2 \frac{\sqrt{3}\, s_W^2\, X}{\sqrt{(1-4s_W^2)}} \; ; \; a_Z^{/f} = - \sqrt{(1-4s_W^2)}\, T_{8L}$$

**TABLE 2**

$Z \to f \bar{f}$ and $Z^/ \to f \bar{f}$ couplings

| Fermion | ChargeQ | $v_Z^f$ | $a_Z^f$ | $v_Z^{/f}$ | $a_Z^{/f}$ |
|---|---|---|---|---|---|
| $d_i$ | - 1 / 3 | $-\frac{1}{2}+\frac{2}{3}s_W^2$ | $\frac{1}{2}$ | $\frac{1-8s_W^2}{2\sqrt{3}(1-4s_W^2)}$ | $\frac{-\sqrt{(1-4s_W^2)}}{2\sqrt{3}}$ |
| $U_i$ | +2 / 3 | $\frac{1}{2}-\frac{4}{3}s_W^2$ | $-\frac{1}{2}$ | $\frac{1-8s_W^2}{2\sqrt{3}(1-4s_W^2)}$ | $\frac{-\sqrt{(1-4s_W^2)}}{2\sqrt{3}}$ |
| $D_i$ | - 4/3 | $\frac{8}{3}s_W^2$ | 0 | $\frac{-1+2s_W^2}{2\sqrt{3}(1-4s_W^2)}$ | $\frac{\sqrt{(1-4s_W^2)}}{\sqrt{3}}$ |
| t | + 2/3 | $\frac{1}{2}-\frac{4}{3}s_W^2$ | $-\frac{1}{2}$ | $\frac{1+4S_W^2}{2\sqrt{3}(1-4s_W^2)}$ | $\frac{-\sqrt{(1-4s_W^2)}}{2\sqrt{3}}$ |
| B | -1/3 | $-\frac{1}{2}+\frac{2}{3}s_W^2$ | $\frac{1}{2}$ | $\frac{1+4S_W^2}{2\sqrt{3}(1-4s_W^2)}$ | $\frac{-\sqrt{(1-4s_W^2)}}{2\sqrt{3}}$ |
| T | 5/3 | $-\frac{10}{3}s_W^2$ | 0 | $\frac{-1+8s_W^2}{2\sqrt{3}(1-4s_W^2)}$ | $\frac{\sqrt{(1-4s_W^2)}}{\sqrt{3}}$ |
| ν | 0 | $\frac{1}{2}$ | $-\frac{1}{2}$ | $\frac{\sqrt{(1-4s_W^2)}}{2\sqrt{3}}$ | $\frac{-\sqrt{(1-4s_W^2)}}{2\sqrt{3}}$ |
| $e^-$ | -1 | $-\frac{1}{2}+2s_W^2$ | $\frac{1}{2}$ | $\frac{\sqrt{(1-4s_W^2)}}{2\sqrt{3}}$ | $-\frac{\sqrt{(1-4s_W^2)}}{2\sqrt{3}}$ |
| $P^+$ | +1 | $-2s_W^2$ | 0 | $-\frac{\sqrt{(1-4s_W^2)}}{\sqrt{3}}$ | $\frac{\sqrt{(1-4s_W^2)}}{\sqrt{3}}$ |

$$v_Z^{f} = g_L + g_R; \quad a_Z^{f} = -g_L + g_R$$

$$v_Z^{/f} = g_L^{/} + g_R^{/}; \quad a_Z^{/f} = -g_L^{/} + g_R^{/}$$

The Lagrangian for the fourth neutral gauge boson is given by

$$L(Z^{//}) = \frac{2 g\, s_W\, Z^{//\mu}}{\sin 2\theta\ \sqrt{(1-4s_W^2)}} \Sigma_f\, \bar{f}\, \gamma_\mu [(\sqrt{6}T_{15} - \sin^2\theta\, X)P_L + (-\sin^2\theta\, X)P_R] f$$

$$= \frac{2 g\, s_W\, Z^{//\mu}}{\sin 2\theta\ \sqrt{(1-4s_W^2)}} \Sigma_f\, \bar{f}\, \gamma_\mu (g_L P_L + g_R P_R) f$$

$$= \frac{g\, s_W\, Z^{//\mu}}{\sin 2\theta\ \sqrt{(1-4s_W^2)}} \Sigma_f\, \bar{f}\, \gamma_\mu (v_Z^{//f} + a_Z^{//f}\gamma_5) f \qquad (32)$$

where $v_Z^{//f} = g_L + g_R = \sqrt{6}T_{15} - 2\sin^2\theta\, X; \quad a_Z^{//f} = -g_L + g_R = -\sqrt{6}T_{15}$

**5(b) Charged currents**

The Charged Current Lagrangian is given by

$$L^{CC}_{lepton} = \frac{g}{\sqrt{2}} [\, \bar{\nu}_\alpha \gamma_\mu (e_\alpha^- W^{+\mu} + P^+_\alpha Y^{\mu}) + \bar{e}_\alpha^- \gamma_\mu (\nu_\alpha W^{-\mu} + P^+_\alpha Y^{--\mu})$$

$$+ \bar{P}^+_\alpha \gamma_\mu (\nu_\alpha Y^{+\mu} + e_\alpha^- Y^{\mu ++})] + H.c. \qquad (33)$$

$$L^{CC}_{quark} = -\frac{g}{\sqrt{2}} [\, \bar{d}_i \gamma_\mu (u_i W^{-\mu} + D_j Y^{+\mu}) + \bar{u}_i \gamma_\mu (W^{+\mu} d_i + Y^{++\mu} D_j)$$

$$+ \bar{D}_j \gamma_\mu (d_i Y^{-\mu} + u_i Y^{-\mu}) + \bar{t} \gamma_\mu (b W^{+\mu} + T Y^{\mu})$$

$$+ \bar{b} \gamma_\mu (t W^{-\mu} + T Y^{-\mu}) + \bar{T} \gamma_\mu (t Y^{+\mu} + b Y^{++\mu})] + H.c.$$

where $\alpha = 1,2,3$ ; i = 1,2,3 and j = 1,2 generation indices. (34)

**5(c) Flavour –changing Neutral Currents (FCNC)**

The FCNC processes in 3-3-1 models have been analysed recently[15] for several cases.We consider the coupling to fermions as mediated only by $Z^{/}$ gauge boson for a small mixing angle for Z , $Z^{/}$ .The FCNC Lagrangian mediated by Z, $Z^{/}$ contribute in left-handed sector only and are a result of different X quantum numbers for the third and first two generations .

$$L_{FCNC} = \frac{-g}{C_W} J_\mu(Z')Z'^\mu - \frac{2 g s_W J_\mu(Z'')Z''^\mu}{\sin 2\theta\sqrt{(1-4s_W^2)}} \quad (35)$$

$$J_\mu(Z') = \Sigma_{i,j} \bar{f_i} \gamma_\mu (g_L')_{ij} P_L f_j \;;\; J_\mu(Z'') = \Sigma_{i,j} \bar{f_i} \gamma_\mu (g_L'')_{ij} P_L f_j$$

$$g_L' = \frac{\sqrt{3}\, s_W^2 X.}{\sqrt{(1-4s_W^2)}} \;;\; g_L'' = -\sin^2\theta X \quad (36)$$

The mass eigenstates $f_i$ can be related to gauge states [16] $Ui'$( i = u, c ,t ), $D_i'$(i = d, s, b) for up and down sectors by unitary matrices $V_{ij}$, $W_{ij}$

$$f_i^{\,u} = V_{ij} U_j' ;\; f_i^{\,d} = W ij D_j'$$

- For up-sector FCNC,

$$J_\mu^u (Z') = \Sigma_{i,j} V_{3i}^* U_i' \gamma_\mu (g_L')_{ij} P_L V_{3j} U_j' = \Sigma_{i,j} \bar{U}_i' \gamma_\mu (B^u)_{ij} P_L U_j'$$

where $B^u{}_{ij} = V_{3i}^*(g_L')_{ij} V_{3j}$ . (37)

- For down sector FCNC

$$J_\mu(Z') = \Sigma_{i,j} W_{3i}^* \bar{D}_i' \gamma_\mu (g_L')_{ij} P_L W_{3j} D_j' = \Sigma_{i,j} \bar{D}_i' \gamma_\mu (B^d)_{ij} P_L D_j'$$

$$B^d{}_{ij} = W_{3i}^*(g_L')_{ij} W_{3j} . \quad (38)$$

The FCNC Lagrangian for $Z'$ transitions

$$L_{FCNC} = \frac{-g}{c_W} (\sin\varphi Z_1^\mu + \cos\varphi Z_2^\mu) \sum_{i,j} [\, \bar{U}_i B^u{}_{ij} \gamma_\mu P_L U_j + D_i \gamma_\mu B^d{}_{ij} P_L D_j] \quad (39)$$

**5(c) Fermion Masses and mixing**

The Yukawa interactions in 3-3-1 model include additional terms for new charged quarks and Higgs scalars. From Table 1 we obtain Yukawa couplings and Lagrangian

$$L_Y = \sum_k [\, h^k{}_t (a^*M^*)(ac)(c^*M)_k + h_b{}^k (a^*M^*)(ac^*)(cM)_k ] + h_T(a^*M^*)(ab)(b^*M)$$

$$+ \sum_i (a M^*)_i [\sum_k \{ h_u{}^{ik}(a^*c)(c^*M)_k + h_d{}^{ik}(a^*c^*)(cM)_i \} + (a^*b^*) \sum_j h_D{}^{ij}(bM)_j ]$$

where k = 1,2,3 and i,j = 1,2. (40)

- The Yukawa Lagrangian for quarks

$$L_Y^{\,Q} = \bar{Q}_{3L} \sum_k [\, \eta\, h_t^{\,k} u_{kR} + \rho\, h_b^{\,k} d_{kR} \,] + \bar{Q}_{3L} \chi\, h_T T_R$$

$$+ \sum_i \bar{Q}_{iL} \, [ \sum_k \{ \rho^* h_u^{\,ik} u_{kR} + \eta^* h_d^{\,ik} d_{kR} \} + \chi^* \sum_j h_D^{\,ij} D_{jR} ] \quad (41)$$

where h 's are Yukawa coupling constants with k = 1,2,3 and i ,j = .1,2

From above we get the mass matrix for 2/3 and – 1/3 charged quarks of the form

$$M_u = \begin{pmatrix} h_u^{11} u & h_u^{12} u & h_u^{13} u \\ h_u^{21} u & h_u^{22} u & h_u^{23} u \\ h_t^{1} v & h_t^{2} v & h t^{3} v \end{pmatrix} \quad (42\,)$$

$$M_d = \begin{pmatrix} h_d^{11} v & h_d^{12} v & h_d^{13} v \\ h_d^{21} v & h_d^{22} v & h_d^{23} v \\ h_b^{1} u & h_b^{2} u & h_b^{3} u \end{pmatrix} \quad (43)$$

For exotic charged quark T, the mass $m_T = h_T\, V$ while for $D_i$ ( i = 1,2 )

$$M_D = \begin{pmatrix} h_D^{11} V & h_D^{12]} V \\ h_D^{21} V & h_D^{22} V \end{pmatrix} \quad (44)$$

The $D_1$, $D_2$ quarks also acquire masses proportional to V .

- The lepton Yukawa couplings are

$$L^Y_l = (\, a\, c\, ^*) \sum_{k,m} h_{km} \, (a\, ^*N^*)_k \, (c\, N)_m + (a\, b) \sum_{k,m} h_{km} \, (a^* N^*)_k \, (b^*N)_m \quad (45)$$

where k,m = 1,2,3.

The mass terms for lepton are

$$L_l^M = -\sum_{k,m} [h^l_{km} \bar{\psi}_{Lk} \psi_{Rm} \rho + h^P_{km} \bar{P}_{Lk} P_{Rm} \chi] \quad (46)$$

where k , m = 1,2,3 family indices .The lepton field $\psi_{Lk} = (e, \mu, \tau)_L$ while the right-handed $\Psi_{Rm} = (e_R, \mu_R, \tau_R)$ The heavy leptons include $P_{Lk} = (P_1, P_2, P_3)_L$ and right-handed $P_{Rm} = (P_1, P_2, P_3)_R$ The interaction terms are

$$L_l^{int} = -\sum_{k,m} [h'_{km} \bar{\psi}_{Lk} P_{Rm} \chi + h'_{km} \bar{P}_{Lk} \psi_{Rm} \rho] \quad (47)$$

The right-handed neutrino is a singlet (1,1,0) in the model and optional since it does not affect anomaly cancellations

**Quark Masses :** To obtain quark masses we consider mass states $f_i$ to be related to gauge states by $U_i'$ , $Di'$ by unitary matrices $V_{L,R}$ and $W_{L,R}$

$$\begin{pmatrix} u_1 \\ u_2 \\ u_3 \end{pmatrix}_{L,R} = V_{L,R} \begin{pmatrix} u \\ c \\ t \end{pmatrix}_{L,R} \qquad \begin{pmatrix} d_1 \\ d_2 \\ d_3 \end{pmatrix}_{L,R} = W_{L,R} \begin{pmatrix} d \\ s \\ b \end{pmatrix}_{L,R}$$

We consider the following conditions for Yukawa couplings

- $h_u, h_d << h_t, h_b$
- $h_t \sim h_b = h' = \sqrt{\Sigma_i (h_t^i)^2}$ ; i = 1,2,3

After transforming $M_q M_q^\dagger$ to diagonal form , we obtain masses for quarks at tree level as

$$m_t / m_b = v/u ; \; m_u = m_b = 0, M_D = h_D V, M_T = h_T V \quad (48)$$

**Lepton Masses**: Leptons include three generations of heavy leptons ($P_1$, $P_2$ ,$P_3$) with masses $M_P = h^P V$ . At tree level the charged fermions ( $e^-$, $\mu^-$,$\tau^-$) obtain masses $m_e = \Gamma u$ where

$$\Gamma = \begin{pmatrix} h^{l}_{e} & h_{12} & h_{13} \\ h_{21} & h_{\mu} & h_{23} \\ h_{31} & h_{32} & h_{\tau} \end{pmatrix} \qquad (49)$$

We consider $h_{ij} = 0$ for $i \neq j$ so that only diagonal terms contribute to charged lepton masses.The exotic fermion masses are all proportional to V .For neutrinoes , one-loop diagrams have been suggested for no right-handed $\nu_R^c$ neutrino case [15]

**6.Phenomenology of neutral gauge bosons and $Z^{/}$, $Z^{//}$ decays**

The invariant amplitudes for various decay channels for the neutral gauge bosons can be considered to calculate the decay rates. The decay modes include $Z^{/} \rightarrow f\,\bar{f}$ flavor-conserving and flavor changing cases and $Z^{/} \rightarrow ZH$ where H is a Higgs scalar.The $Z^{/}$ decay channels have been considered in 3-3-1 model without charged heavy leptons and for a = √3case [13].We consider these cases in the present formalism.

**6.1 $Z^{/} \rightarrow f_i\,\bar{f}_i$ decays(flavor conserving case)**

From eqn.(31)the lagrangian

$$L = \frac{-ig}{2c_W} Z^{/\mu} \sum_f \{ \bar{f}\gamma_\mu (v_Z^{/f} + a_Z^{/f}\gamma_5) f \}$$

where $v_Z^{/f}$ and $a_Z^{/f}$ are vector and axial currents as in Table 2.

The amplitude for fermion f (quarks and leptons) is

$$M = \frac{-g}{2c_W} \varepsilon_\mu(p)\ \bar{u}(l)\,\gamma^\mu\,(v_Z^{/f} + a_Z^{/f}\gamma_5)\,v(k) \qquad (50)$$

From Feynman rules, the decay rate

$$\Gamma = \alpha \frac{M_{Z^/} N_c}{3\, s_{2W}^{\ 2}} [\,|v_Z^{/f}|^2\{1 + \frac{2\, m_f^{\ 2}}{M_Z^{/2}}\} + |a_Z^{/f}|^2\{1 - \frac{4\, m_f^{\ 2}}{M_Z^{/2}}\}]\ (1 - 4\frac{m_f^{\ 2}}{M_Z^{/2}})^{1/2} \qquad (51)$$

where $\alpha$ is fine structure constant . $\alpha = \frac{e^2}{4\pi}$. For quarks (lepton) , $N_c = 3$ (1)

From Table 2, the $Z^/ \rightarrow e^+ e^-, \nu\ \bar{\nu}, P^+ P^-$ decays are proportional to $\sqrt{(1-4s_W^2)}$ and suppressed giving a leptophobic nature to $Z^/$. For $Z^/ \rightarrow t\ \bar{t}, b\ \bar{b}$ decays, only vector current $v_Z^{/f}$ contributes significantly .

$$\Gamma(Z^/ \rightarrow t\ \bar{t}) = \frac{\alpha M_{Z^/} N_c}{3\, s_{2W}^2} |v_Z^{/f}|^2 \left(1 + \frac{2\, m_t^2}{M_{Z^/}^2}\right)\left(1 - 4\frac{m_t^2}{M_{Z^/}^2}\right)^{1/2} \qquad (52)$$

For $s_W^2 = 0.23$, $v_Z^{/t} = \frac{1 + 4s_W^2}{2\sqrt{3}(1-4s_W^2)} = 1.9595$; <v> = 224GeV

< V> = 1 TeV, $M_Z^/ = 1.13$ TeV; $\Gamma(Z^/ \rightarrow t\ \bar{t}) = 0.0446\ TeV.$

<V> = 2 TeV, $M_Z^/ = 2.26$ TeV; $\Gamma(Z^/ \rightarrow t\ \bar{t}) = 0.0895TeV$

Since $M_D$, $M_T$ exotic quark masses are proportional to <V> , $M_Z^/ < 2M_{D,T}$ so that $Z^/$ cannot decay to $\bar{D}D$, $\bar{T}\,T$ exotic quarks. This also applies to the decay of $Z^/$ to $\bar{P}P^+$ charged heavy leptons.

**6.2 $Z^/ \rightarrow f_i\ \bar{f_j}$ flavor changing decays**.

The amplitude for $Z^/$ decay to t and q (c or u)

$$M(Z^/ \rightarrow \bar{U}^/_i U^/_j) = \frac{-g\varepsilon^\mu (Z^/)}{C_W} \bar{U}^/_i \gamma_\mu B^u_{ij} P_L U^/_j \qquad (52)$$

$$For\ Z^/ \rightarrow q\ \bar{t}, \qquad B^u_{tc} = -\frac{|V^*_{tq} V_{tt}|\sqrt{3}s_W^2}{\sqrt{(1-4\,s_W^2)}},$$

$$\Gamma(Z^/ \rightarrow q\ \bar{t}) = \frac{4\alpha M_Z^/ s_W^4 N_c |V^*_{tq} V_{tt}|^2}{s^2_{2W}(1-4s_W^2)}\left[1 - \left\{\frac{(m_t^2 + m_q^2)}{2M_Z^{/2}} + \frac{(m_t^2 - m_q^2)^2}{M_Z^{/4}}\right\}\right]$$

$$\cdot\sqrt{\left\{1 + \frac{(m_t^2 - m_q^2)^2}{M_Z^{/4}} - 2\frac{(m_t^2 + m_q^2)}{M_Z^{/2}}\right\}} \qquad (53)$$

The $|V^*_{tq} V_{tt}|$ factor obtained from $Z^/$ t c couplings are proportional to products of Kobayashi –Masakawa matrix elements and $|V^*_{tq} V_{tt}|^2 = 1.9 \times 10^{-3}$ ;

For the decay $Z^/ \rightarrow c\ \bar{t}$ [13]

At <V> = 1 TeV, $M_Z^{/}$ = 1.13 TeV, $\Gamma( Z^{/} \rightarrow c\ \bar{t})$ = 0.389 x $10^{-3}$ TeV.

### 6.3 $Z^{/} \rightarrow Z\ H_1^0$ decay mode

The decay of $Z_2 \rightarrow Z_1 H^0$ can be considered for small Z-$Z^{/}$ mixing as $Z^{/}$ to ZH mode,

$$M = \varepsilon^{\mu}(k^{/}) \frac{(-g^2 v)}{\sqrt{(u^2+v^2)}} g_{\mu\nu} \varepsilon^{\nu *}(k) \tag{54}$$

$$\Gamma(Z^{/} \rightarrow ZH^0) = \frac{\pi\alpha^2 v^2 \sin^2\beta}{3 s_W^4 M_Z^{/}} \left[ -1 + \frac{1}{4M_Z^2}\left\{ 1 + \frac{1}{M^4{}_Z^{/}}(M^2{}_Z - M^2{}_H)^2 \right\}\right]\left[ 1 - \frac{2}{M^2{}_Z^{/}}(M^2{}_H + M_Z^2) + \frac{1}{M_Z^{/4}}(M_H^2 - M_Z^2)^2 \right]^{1/2} \tag{55}$$

These decay modes can contribute to single Higgs production in $Z^{/} \rightarrow t\ \bar{t}\ H^0$ decays.

**7.Results and Conclusions**

The basic motivation of this work is to extend the 3-3-1 gauge symmetry to 3-3-1-1 which can be embedded in $SU(4)_C \otimes SU(4)_{L+R}$ gauge group. The special feature of the model is that B-L number is well defined so that the gauge bosons are not bileptons .The charged heavy lepton version of 3-3-1 model with two possible electric charge assignments is obtained[8] .An interesting bifundamental picture emerges for fermions and scalars as a consequence of both (3,4,1) and $(3,4^*,1)$ representations for three generation anomally –free fermions of the 3-3-1 model. The gauge boson sector is extended by an additional $Z^{//}$ which decouples from the other three neutral gauge fields. However, the $Z^{//}$ couples to ordinary fermions .We consider a pattern of symmetry breaking with $SU(3)_L \rightarrow SU(2)_L \otimes U(1)_{X^/}$ where $X^/ = \sqrt{3}T_{8L}$ and hypercharge

$Y = 2(X - X^/)$. The masses for exotic charged ( D,T) quarks and heavy leptons P are proportional to VEV of the scalar $\chi$ , $< \chi > = V \sim 1$ TeV. Only top and bottom quarks get masses at tree level while the masses of ordinary fermions cannot be obtained at tree level. These have been generated by one-loop diagrams in a SUSY formalism for

$a = 1/\sqrt{3}$ 3-3-1 model [17,18]

We present some phenomenological consequences of the model for neutral gauge boson decays. The $Z^/$ decays are obtained at tree level with quark, lepton pairs and single Higgs in the final states. The masses of $Z^/$, $P^+$, $D_{1,2}$, T exotics depend on VEV of scalar $\chi^0$ , (<V>) and exotic decay modes are restricted ( $M_{Z^/} > 2M_f$ ) for $Z^/$ gauge boson in the present model. The leptonic modes are suppressed due to a factor $\sqrt{(1-4s_W^2)}$ where $s_W^2 = 0.23$ thus predicting a leptophobic nature for $Z^/$.The model is a straightforward 3-3-1 extension of the Standard Model symmetry and offers interesting phenomenology for new physics at TeV scale.A supersymmetric version of the model with extended Higgs sector can lead to interesting phenomenology for leptons and one –loop quantum effects